\documentclass[aps,prd,twocolumn,showpacs,floatfix]{revtex4}

\usepackage{amsmath}
\usepackage{amssymb}
\usepackage{graphicx}
\usepackage{bm}

\begin{document}
\title{Conformally symmetric traversable wormholes}

\author{Christian G. B\"ohmer}
\email{christian.boehmer@port.ac.uk}
\affiliation{Institute of Cosmology \& Gravitation,
             University of Portsmouth, Portsmouth PO1 2EG, UK}

\author{Tiberiu Harko}
\email{harko@hkucc.hku.hk}
\affiliation{Department of Physics and Center for Theoretical
             and Computational Physics, The University of Hong Kong,
             Pok Fu Lam Road, Hong Kong}

\author{Francisco S.~N.~Lobo}
\email{francisco.lobo@port.ac.uk}
\affiliation{Institute of Cosmology \& Gravitation,
             University of Portsmouth, Portsmouth PO1 2EG, UK}
\affiliation{Centro de Astronomia e Astrof\'{\i}sica da
             Universidade de Lisboa, Campo Grande, Ed. C8 1749-016 Lisboa,
             Portugal}

\date{\today}

\begin{abstract}
Exact solutions of traversable wormholes are found under the
assumption of spherical symmetry and the existence of a {\it
non-static} conformal symmetry, which presents a more systematic
approach in searching for exact wormhole solutions. In this work,
a wide variety of solutions are deduced by considering choices for
the form function, a specific linear equation of state relating
the energy density and the pressure anisotropy, and various
phantom wormhole geometries are explored. A large class of
solutions impose that the spatial distribution of the exotic
matter is restricted to the throat neighborhood, with a cut-off of
the stress-energy tensor at a finite junction interface, although
asymptotically flat exact solutions are also found. Using the
``volume integral quantifier,'' it is found that the conformally
symmetric phantom wormhole geometries may, in principle, be
constructed by infinitesimally small amounts of averaged null
energy condition violating matter. Considering the tidal
acceleration traversability conditions for the phantom wormhole
geometry, specific wormhole dimensions and the traversal velocity
are also deduced.
\end{abstract}

\maketitle

\section{Introduction}

Traversable wormholes are hypothetical tunnels in space and time
\cite{Morris:1988cz}, permitting effective `superluminal travel',
although the speed of light is not {\it locally}
surpassed~\cite{Lobo:2004wq}, and induce closed timelike curves
\cite{Morris:1988tu}, apparently violating causality. These
geometries are supported by material, denoted as {\it exotic
matter}, that violates the null energy condition. In fact, they
violate all the known pointwise energy conditions and averaged
energy conditions, which are fundamental to the singularity
theorems and theorems of classical black hole thermodynamics
\cite{Visser}. Although classical forms of matter are believed to
obey these energy conditions~\cite{hawkingellis}, it is a
well-known fact that they are violated by certain quantum fields,
amongst which we may refer to the Casimir effect. The literature
is rather extensive in candidates for wormhole spacetimes, and one
may mention several cases, ranging from scalar fields
\cite{WHscalar}, wormhole geometries in higher dimensions
\cite{Kar,BraneWH}, spacetimes in Brans-Dicke theory \cite{Nandi},
solutions supported by semi-classical gravity (see Ref.
\cite{Garattini:2007ff} and references therein), geometries in the
context of nonlinear electrodynamics~\cite{Arellano}, to wormholes
supported by equations of state responsible for the accelerated
expansion of the universe
\cite{phantomWH,phantomWH2,EOSaccel,gonzalez}, etc. We refer the
reader to Refs. \cite{Visser,Lemos:2003jb} and references therein
for more details.

Essentially, wormhole physics is a specific example of adopting
the reverse philosophy of solving the Einstein field equations, by
first constructing the spacetime metric, then deducing the
stress-energy tensor components. It is interesting to consider a
more systematic approach in searching for exact solutions. For
instance, one may adopt the approach outlined in Refs.
\cite{Herrera,Maartens:1989ay}, where the spacetime is assumed to
be spherically symmetric and to possess a conformal symmetry. If
the vector ${\bf \xi}$ generates this conformal symmetry, then the
metric ${\bf g}$ is conformally mapped onto itself along ${\bf
\xi}$. This is translated by the following relationship
\begin{equation}
{\cal L}_{\bf \xi} \,{\bf g}=\psi\, {\bf g} \,,
   \label{Lieconf}
\end{equation}
where ${\cal L}$ is the Lie derivative operator and $\psi$ is the
conformal factor.

As emphasized in Ref.~\cite{Maartens:1989ay}, despite being
essentially geometric in character, this approach is physically
justifiable, namely, it is a generalization of self-similarity in
hydrodynamics; and it also generalizes the property of the
incompressible Schwarzschild interior solution which is conformally
flat and therefore is characterized by an additional symmetry.
Indeed, if the energy density is non-constant the spacetime is
no longer conformally flat, see e.~g.~\cite{confpf}.
The above conformal symmetry has been of particular interest in
the context of static and spherically symmetric perfect fluid solutions.
In general, an equation of state needs to be specified to close the
system of differential equations, this choice, however, is an
essentially non-geometric one. On the other hand, the conformal
symmetry can be regarded as a geometrical equation of state
closing the system of equations.

The approach outlined in Ref.~\cite{Herrera} considers a static conformal
symmetry, i.e., a static $\xi$, which is essentially responsible
for the singular solutions at the stellar centers found. For this
reason, non-static conformal symmetries, i.e., non-static $\xi$
and static $\psi$, were considered in Ref. \cite{Maartens:1989ay},
and we shall essentially follow this approach in this work, as it
provides a wider range of exact wormhole solutions. We do,
however, emphasize that the singular character of the solutions at
the stellar center in Ref. \cite{Herrera} need not be problematic
in wormhole physics, due to the absence of a center, as the radial
coordinate possesses a minimum value, $r_0$, denoted as the
wormhole throat. Thus, we shall also briefly analyze the static
$\xi$ solution, namely, in the phantom wormhole context.
It is interesting to note that an
exact analytical solution describing the interior of a charged
strange quark star was found~ \cite{Mak:2003kw}; solutions were
also explored in brane-worlds~\cite{Harko:2004ui}; and also in the
context of the galactic rotation curves~\cite{Mak:2004hv}.

This paper is outlined in the following manner: In Section
\ref{Sec1:field}, the field equations are presented; in Section
\ref{Sec2:conformal}, exact general solutions are deduced using
non-static conformal symmetries; in Section \ref{Sec3:solutions},
a wide variety of solutions are deduced by considering choices for
the form function, a specific linear equation of state relating
the energy density and the pressure anisotropy, and various
phantom wormhole geometries are explored; and in Section
\ref{Sec4:Conclusion} we conclude.

\section{Field equations}\label{Sec1:field}

The spacetime metric representing a spherically symmetric and
static wormhole is given by
\begin{equation}
ds^2=-e ^{2\Phi(r)}\,dt^2+\frac{dr^2}{1- b(r)/r}+
r^2 d\Omega^2 \label{metricwormhole}\,,
\end{equation}
where $\Phi(r)$ and $b(r)$ are arbitrary functions of the radial
coordinate, $r$, denoted as the redshift function, and the form
function, respectively \cite{Morris:1988cz}. The radial coordinate
has a range that increases from a minimum value at $r_0$,
corresponding to the wormhole throat, to $a$, where the interior
spacetime will be joined to an exterior vacuum solution. Specific
asymptotically flat wormhole geometries will also be considered,
where $r$ extends from the throat out to infinity.

To be a wormhole solution, several conditions need to be imposed
\cite{Morris:1988cz}: First, one needs to verify the absence of
event horizons, so that the redshift function $\Phi(r)$ is finite
throughout the range of interest; In second place, the mathematics
of embedding imposes a flaring-out condition, translated by the
following condition, $(b'r-b)/b^2<0$, which reduces to $b'<1$ at
the throat, and taking into account the field equations, it is
this condition that implies the violation of the null energy
condition, as shown below; The conditions $(1-b/r)>0$ and
$b(r_0)=r_0$ at the throat are also imposed.

Using the Einstein field equation, $G_{\mu\nu}=\kappa^2
\,T_{\mu\nu}$ (with $\kappa^2=8\pi$ and $c=G=1$), we obtain the
following non-zero stress-energy tensor components
\begin{eqnarray}
\rho(r)&=&\frac{1}{\kappa^2} \;\frac{b'}{r^2}   \label{rhoWH}\,,\\
p_r(r)&=&\frac{1}{\kappa^2} \left[2 \left(1-\frac{b}{r}
\right) \frac{\Phi'}{r} -\frac{b}{r^3}\right]  \label{prWH}\,,\\
p_t(r)&=&\frac{1}{\kappa^2} \left(1-\frac{b}{r}\right)\Bigg[\Phi
''+ (\Phi')^2+\frac{\Phi'}{r}
   \nonumber   \\
&&- \frac{b'r-b}{2r(r-b)}\Phi'-\frac{b'r-b}{2r^2(r-b)} \Bigg]
\label{ptWH}\,,
\end{eqnarray}
where $\rho(r)$ is the energy density, $p_r(r)$ is the radial
pressure, and $p_t(r)$ is the lateral pressure measured in the
orthogonal direction to the radial direction. Note that the
conservation of the stress-energy tensor, $T^{\mu\nu}{}_{;\nu}=0$,
provides the following relationship
\begin{equation}
p_r'=\frac{2}{r}\,(p_t-p_r)-(\rho +p_r)\,\Phi '
\label{prderivative} \,.
\end{equation}

A fundamental property of wormholes is the violation of the null
energy condition (NEC), $T_{\mu\nu}k^\mu k^\nu \geq 0$, where
$k^\mu$ is {\it any} null vector \cite{Morris:1988cz}. From Eqs.
(\ref{rhoWH}) and (\ref{prWH}), one verifies
\begin{equation}
\rho(r)+p_r(r)=\frac{1}{8\pi}\,\left[\frac{b'r-b}{r^3}+
2\left(1-\frac{b}{r}\right) \frac{\Phi '}{r} \right]  \,.
        \label{NECviol}
\end{equation}
Taking into account the flaring-out condition and the finite
character of $\Phi(r)$, evaluated at the throat $r_0$, we have
$\rho+p_r<0$. Matter that violates the NEC is denoted {\it exotic
matter}. More specifically, in terms of the form function
evaluated at the throat, we have $b'(r_0)<1$.

\section{Conformal symmetry}\label{Sec2:conformal}

Applying a systematic approach in order to deduce exact solutions,
we shall take into account the method used in Ref.
\cite{Maartens:1989ay}, where the static and spherically symmetric
spacetime possesses a non-static conformal symmetry. It should be
emphasized that neither $\xi$ nor $\psi$ need to be static even
though one considers a static metric. Note that Eq. (\ref{Lieconf})
takes the following form
\begin{equation}
g_{\mu\nu,\alpha}\,\xi^\alpha+g_{\alpha\nu}\,\xi^{\alpha}{}_{,\mu}
+g_{\mu\alpha}\,\xi^{\alpha}{}_{,\nu}=\psi \,g_{\mu\nu}   \,.
  \label{Lie}
\end{equation}
We shall follow closely the assumptions made in Ref.
\cite{Maartens:1989ay}, where the condition
\begin{equation}
\mathbf{\xi}=\alpha(t,r)\,\partial_t+\beta(t,r)\,\partial_r\,,
    \label{Lie3}
\end{equation}
is considered, and the conformal factor is static, i.e.,
$\psi=\psi(r)$.

Taking into account metric (\ref{metricwormhole}), then Eq.
(\ref{Lie}) provides the following solutions
\begin{eqnarray}
\alpha=A+\frac{kt}{2}, \qquad
\beta=\frac{1}{2}Br\sqrt{1-\frac{b(r)}{r}} \,
    \label{Liesolutin1}
\end{eqnarray}
and
\begin{eqnarray}
\psi(r)&=&B\sqrt{1-\frac{b(r)}{r}} \,,
    \label{Liesolutin2a} \\
e^{2\Phi(r)}&=&C^2r^2\exp\left(-\frac{2k}{B}\int
\frac{dr}{r\sqrt{1-\frac{b(r)}{r}}}\right) \,,
   \label{Liesolutin2}
\end{eqnarray}
where $A$, $B$, $C$ and $k$ are constants. Note that, without a
loss of generality, one may consider $A=0$ as $A\partial_t$ is a
Killing vector, and $B=1$ by rescaling ${\bf \xi}$ and $\psi$ in
the following manner: $\xi \rightarrow B^{-1}\xi$ and $\psi
\rightarrow B^{-1}\psi$, which leaves Eq. (\ref{Lie}) invariant.
Thus, Eqs. (\ref{Lie3}) and (\ref{Liesolutin1}) reduce to
\begin{eqnarray}
\xi=\frac{1}{2}kt \,\partial_t+\frac{1}{2}\psi(r)r\,\partial_r \,
    \label{Liesolutin1b}
\end{eqnarray}
and Eqs. (\ref{Liesolutin2a})--(\ref{Liesolutin2}) take the form
\begin{eqnarray}
b(r)&=&r[1-\psi^2(r)] \,,  \label{Liesolutin2b} \\
\Phi(r)&=&\frac{1}{2}\ln(C^2r^2)-k\int
\frac{dr}{r\sqrt{1-\frac{b(r)}{r}}} \,.
   \label{Liesolutin2c}
\end{eqnarray}
An interesting feature of these solutions that immediately stands
out, by taking into account Eq. (\ref{Liesolutin2b}), is that the
conformal factor is zero at the throat, i.e., $\psi(r_0)=0$.

The existence of conformal motions imposes strong constraints on
the wormhole geometry, so that the stress-energy tensor components
are written solely in terms of the conformal function, and take
the following form
\begin{eqnarray}
\rho(r)&=&\frac{1}{\kappa^2r^2}\left(1-\psi^2-2r\psi\psi' \right)
       \label{rhoWH2}\,,\\
p_r(r)&=&\frac{1}{\kappa^2r^2} \left(3\psi^2-2k\psi-1\right)
       \label{prWH2}\,,\\
p_t(r)&=&\frac{1}{\kappa^2r^2}\left(\psi^2-2k\psi+k^2+2r\psi\psi'\right)
       \label{ptWH2}\,.
\end{eqnarray}

The NEC violation, Eq. (\ref{NECviol}), for this case is given by
\begin{equation}
\rho(r)+p_r(r)=\frac{1}{\kappa^2r^2}\,\left[2\psi(\psi-k)-r(\psi^2)'
\right] \,,
        \label{NECviol2}
\end{equation}
which evaluated at the throat imposes the following condition
$(\psi^2)'>0$.

Several solutions analyzed in this work are not asymptotically
flat, so that one needs to match these interior geometries to an
exterior vacuum spacetime. The spatial distribution of the exotic
matter is restricted to the throat neighborhood, so that the
dimensions of these wormholes are not arbitrarily large. For
simplicity, consider that the exterior vacuum solution is the
Schwarzschild spacetime, given by the following metric
\begin{equation}
ds^2=-\left(1-\frac{2M}{r}\right)\,dt^2+\frac{dr^2}{1-2M/r}+
r^2 d\Omega^2
\label{Sch-metric}.
\end{equation}
Note that the matching occurs at a radius greater than the event
horizon $r_b=2M$, i.e., $a>2M$. The Darmois-Israel formalism
\cite{Darmois-Israel} then provides the following expressions for
the surface stresses of a dynamic thin shell
\cite{phantomWH,LoboCrawford}
\begin{eqnarray}
\sigma&=&-\frac{2}{\kappa^2 a}
\left(\sqrt{1-\frac{2M}{a}+\dot{a}^2}- \sqrt{\psi^2(a)+\dot{a}^2}
\, \right)
    \label{surfenergy}   ,\\
{\cal P}&=&\frac{1}{\kappa^2 a} \Bigg[\frac{1-\frac{M}{a}
+\dot{a}^2+a\ddot{a}}{\sqrt{1-\frac{2M}{a}+\dot{a}^2}}
   \nonumber    \\
&-&\frac{\left(2+\frac{k}{\psi(a)}\right)
\left(\psi^2(a)+\dot{a}^2
\right)+a\ddot{a}-\frac{\psi'(a)a\dot{a}^2}{\psi(a)}}{\sqrt{\psi^2(a)+\dot{a}^2}}
\, \Bigg]         \,,
    \label{surfpressure}
\end{eqnarray}
where the overdot denotes a derivative with respect to the proper
time, $\tau$. $\sigma$ and ${\cal P}$ are the surface energy
density and the tangential surface pressure, respectively. The
static case is given by taking into account $\dot{a}=\ddot{a}=0$
\cite{wormhole-shell}.

\section{Specific solutions}\label{Sec3:solutions}

\subsection{Specific form functions}

In this section, we explore a wide variety of wormhole geometries
by considering specific form functions.

\subsubsection{$b(r)=r_0$}

A particularly interesting case is considering a zero energy
density, which implies a constant form function. Considering
$b(r)=r_0$, and taking into account Eq. (\ref{Liesolutin2c}), we
have
\begin{equation}
e^{2\Phi(r)}=C^2r^2\left(r-\frac{r_0}{2}+r\sqrt{1-\frac{r_0}{r}}\right)^{-2k}
\,.
\end{equation}
An interesting feature of this geometry is the specific case of
$k=1$, which reflects an asymptotically flat spacetime, by
normalizing the constant $C^2=2$, i.e., $e^{2\Phi(r)}\rightarrow
1$ as $r \rightarrow +\infty$. For $k\neq 1$, we need to match
this interior wormhole geometry to an exterior vacuum spacetime at
a junction interface, where the surface stresses are provided by
Eqs. (\ref{surfenergy})-(\ref{surfpressure}).

The stress-energy tensor components are given by, $\rho=0$, and
\begin{eqnarray}
p_r&=&-\frac{1}{\kappa^2}\left(\frac{3r_0-2r}{r^3}+\frac{2k}{r^2}
\sqrt{1-\frac{r_0}{r}}\right) \,, \\
p_t&=&\frac{1}{\kappa^2}\left(\frac{1+k^2}{r^2}-\frac{2k}{r^2}
\sqrt{1-\frac{r_0}{r}}\right) \,,
\end{eqnarray}
where $p_r\rightarrow 0$ and $p_t\rightarrow 0$ as $r \rightarrow
+\infty$.

\subsubsection{$b(r)=r_0^2/r$}

Consider the case of $b(r)=r_0^2/r$, which implies a negative
energy density. From Eq. (\ref{Liesolutin2c}) we have
\begin{equation}
e^{2\Phi(r)}=C^2r^2\left(r+\sqrt{r^2-r_0^2}\right)^{-2k}\,.
\end{equation}
Note that as in the previous example, an asymptotically flat
spacetime is found by considering $k=1$ and $C^2=2$.

The stress-energy tensor components are given by
\begin{eqnarray}
\rho&=&-\frac{1}{\kappa^2}\frac{r_0^2}{r^4} \,, \\
p_r&=&-\frac{1}{\kappa^2}\left(\frac{3r_0^2-2r^2}{r^4}+\frac{2k}{r^2}
\sqrt{1-\frac{r_0^2}{r^2}}\right) \,, \\
p_t&=&\frac{1}{\kappa^2}\left[\frac{1}{r^2}\left(1+k^2+\frac{r_0^2}{r^2}\right)-\frac{2k}{r^2}
\sqrt{1-\frac{r_0^2}{r^2}}\right] \,,
\end{eqnarray}
which tend to zero as $r \rightarrow +\infty$.

\subsubsection{$b(r)=r_0+\gamma^2r_0(1-r_0/r)$}

An interesting form function is $b(r)=r_0+\gamma^2r_0(1-r_0/r)$
\cite{Garattini:2007ff}, with $0<\gamma^2<1$, with provides a
positive energy density. From Eq. (\ref{Liesolutin2c}) we have
\begin{equation}
e^{2\Phi(r)}=C^2r^2\left[r-\frac{r_0}{2}(1+\gamma^2)
+\sqrt{(r-r_0)(r-\gamma^2r_0)}\right]^{-2k} \,,
\end{equation}
and as before, an asymptotically flat spacetime is found by
considering $k=1$ and $C^2=2$.

The stress-energy tensor components are given by
\begin{eqnarray}
\rho&=&\frac{1}{\kappa^2}\frac{\gamma^2r_0^2}{r^4} \,, \\
p_r&=&-\frac{1}{\kappa^2}\Bigg\{\frac{3r_0r(1+\gamma^2)-3\gamma^2r_0^2-2r^2}{r^4}
       \nonumber  \\
&&+\frac{2k}{r^2}
\sqrt{1-\frac{r_0}{r}\left[1+\gamma^2\left(1-\frac{r_0}{r}\right)\right]}\Bigg\} \,, \\
p_t&=&\frac{1}{\kappa^2}\Bigg\{\frac{1}{r^2}\left(1+k^2-\gamma^2\frac{r_0^2}{r^2}\right)
       \nonumber  \\
&&-\frac{2k}{r^2}
\sqrt{1-\frac{r_0}{r}\left[1+\gamma^2\left(1-\frac{r_0}{r}\right)\right]}\Bigg\}
\,,
\end{eqnarray}
which tend to zero as $r \rightarrow +\infty$.

\subsection{Specific equation of state: $\rho=\alpha(p_t-p_r)$}

As was pointed out in the Introduction, wormhole spacetimes are
constructed mainly by designing an appropriate metric and followed
by reconstructing the matter part. An interesting solution is
obtained by assuming that the anisotropy and the energy density
are related by a linear equation of state.  It turns out that an
equation of state of the form $\rho=\alpha(p_t-p_r)$ indeed yields
exact solutions.

In this case the resulting differential equation for the conformal
factor yields a rather complicated expression. However, the form
function admits the following solution
\begin{equation}
b(r) = \frac{1-\alpha(1+k^2)}{(1-2\alpha)}r_0^{\frac{1-2\alpha}{1+\alpha}}
r^{\frac{3\alpha}{(1+\alpha)}}-
\frac{(1-k^2)\alpha}{(1-2\alpha)}r\,.
\end{equation}
It is particularly interesting to note that this form function
satisfies the required wormhole conditions for a wide class of
parameters $\alpha$, which makes this model quite generic. Such
wormhole models, where one assumes the above equation of state for
the matter have not been studied previously.

\subsection{Conformally symmetric phantom wormhole}

An interesting case is that of traversable wormholes supported by
the dark energy equation of state in the phantom regime,
$\omega=p_r/\rho<-1$. Several physical properties and
characteristics have been extensively explored, and we refer the
reader to Refs. \cite{phantomWH,phantomWH2,gonzalez}. Note that
taking into account the field equations (\ref{rhoWH}) and
(\ref{prWH}), the phantom energy equation of state imposes the
following relationship
\begin{equation}
\Phi'(r)=\frac{b+\omega rb'}{2r^2\,\left(1-b/r \right)} \,.
            \label{EOScondition}
\end{equation}
For the case of the non-static conformal factor we could only find
one exact solution in terms of $\psi$, namely if $\omega = -3$,
which is treated in more detail below.

\subsubsection{Static conformal symmetry}

Let us firstly consider
that the components of Eq. (\ref{Lie3}) depend only on the radial
coordinate, $r$. This corresponds to imposing $k=0$ in the
solutions given by Eqs. (\ref{Liesolutin1b}) and
(\ref{Liesolutin2c}). Note that the relationship
(\ref{Liesolutin2b}) still holds, but the expression
(\ref{Liesolutin2c}) now provides the following redshift function
\begin{equation}
e^{2\Phi(r)}=C^2r^2 \,.
            \label{statredshift}
\end{equation}

Thus, the phantom energy differential equation
(\ref{EOScondition}), reduces to
\begin{equation}
\psi'=\frac{(1+\omega)-(3+\omega)\psi^2}{2\omega r\psi} \,.
            \label{EOScondition2}
\end{equation}
The solution for $\psi(r)$ is given by
\begin{equation}
\psi(r)=\pm
\sqrt{Cr^{-(3+\omega)/\omega}+\left(\frac{1+\omega}{3+\omega}\right)}
\label{c1}
\end{equation}
where $C$ is a constant of integration. Note that the solution
given by (\ref{c1}) holds for $\omega \neq 3$ only. This case
must be considered separately, see below.

\begin{figure}[!ht]
\centering
\includegraphics[width=2.8in]{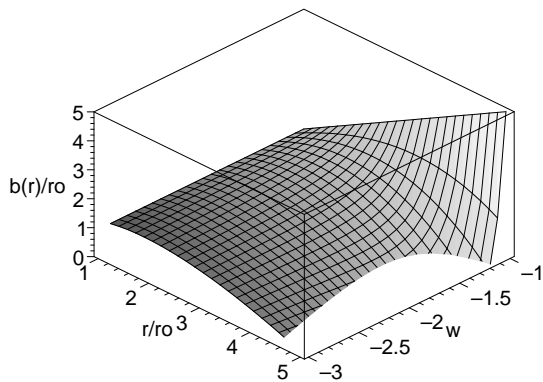}
\caption{The surface represents the adimensional form function
$b(r)/r_0$. Note that to be a wormhole solution one needs to
impose $0<b(r)<r$.} \label{plot1}
\end{figure}

Taking into account Eq. (\ref{Liesolutin2b}), provides the following form
function
\begin{equation}
b(r)=\frac{r_0(1+\omega)}{(3+\omega)}\left(\frac{r}{r_0}\right)^{-3/\omega}
+\frac{2r}{3+\omega} \,.
            \label{phantform}
\end{equation}
The constant of integration $C$ is determined by imposing
$b(r_0)=r_0$. Note that $b'(r_0)=1/|\omega|<1$, which obeys the
flaring out condition at the throat. One also needs to impose
$b(r)>0$ (see Ref. \cite{Lemos:2003jb} for details) and $b(r)<r$,
which is represented by the surface in Fig. \ref{plot1}. Thus, one
needs to match this solution to an exterior spacetime at a
junction interface, $a>2M$.

For $\omega = -3$ the phantom energy differential equation simplifies to
\begin{equation}
\psi'=\frac{1}{3 r\psi} \,,
            \label{c3}
\end{equation}
and yields the following solution for the conformal factor
\begin{equation}
\psi(r)=\pm\sqrt{\frac{2}{3}\log\left(\frac{r}{r_0}\right)} \,.
\label{c2}
\end{equation}
This gives the following form function
\begin{equation}
b(r)=r\left[1-\frac{2}{3}\log\left(\frac{r}{r_0}\right)\right] \,,
\end{equation}
where we have fixed the constant of integration as in the above.
We verify that $b'=1/3$ and since for a wormhole solution $b(r)>0$
has to be imposed, we find that this solutions needs to matched to
an exterior spacetime at some surface $r_0<a < r_0 e^{3/2}$.

\subsubsection{Non-static conformal symmetry}

Let us now consider the specific non-static case where the energy
density and the radial pressure given by~(\ref{rhoWH2})
and~(\ref{prWH2}), respectively, are related by the equation of
state $\omega=-3$. In this case the conformal factor is given by
\begin{equation}
\psi(r) = \frac{1+ {\cal W}(x)}{k}\,, \quad
x = \frac{1}{e}\left(\frac{r_0}{r}\right)^{k^2/3}\,,
\label{n2}
\end{equation}
where ${\cal W}(x)$ is the Lambert W function~\cite{foot1}. As
in the above, the constant of integration was chosen such
that $\psi(r_0)=0$, therefore, at the throat $r_0$ the form
function satisfies $b(r_0)=r_0$. With the above solution
one also verifies that the form function obeys $b'(r_0)<1$.
Moreover, let us also consider the case $k=1$. The limit
$r\rightarrow\infty$ corresponds to $x \rightarrow 0$, which
by definition of the Lambert W function implies ${\cal W}(0) = 0$.
Hence, we conclude that the spacetime is also asymptotically flat
and no further exterior matching is required for this conformal
wormhole solution.

\subsubsection{Volume integral quantifier}

It is also interesting to consider the ``volume integral
quantifier,'' which provides information on the total amount of
matter violating the averaged null energy condition in the
spacetime. This is defined by $I_V=\int[\rho(r)+p_r(r)]dV$ (see
Ref. \cite{Visser:2003yf} for details), and with a cut-off of the
stress-energy at $a$ is given by
\begin{eqnarray}
I_V&=&\left[r\left(1-\frac{b}{r}\right)
\ln\left(\frac{e^{2\Phi}}{1-b/r}\right)\right]_{r_0}^a
    \nonumber    \\
&&-\int_{r_0}^a
\left(1-b'\right)\left[\ln\left(\frac{e^{2\Phi}}{1-b/r}\right)\right]\;dr
    \nonumber   \\
&=&\int_{r_0}^a
\left(r-b\right)\left[\ln\left(\frac{e^{2\Phi}}{1-b/r}\right)\right]'\;dr
\,.
    \label{Iv}
\end{eqnarray}
Taking into account the redshift function, Eq.
(\ref{statredshift}), and the form function, Eq.
(\ref{phantform}), and recalling that $\omega<-1$, one obtains the
following solution for the volume integral quantifier
\begin{equation}
I_V=\left(\frac{1+\omega}{3+\omega}\right)\left[2a-r_0(3+\omega)
+r_0(1+\omega)\left(\frac{a}{r_0}\right)^{-3/\omega}\right] \,.
            \label{sol:Iv}
\end{equation}
Note that for a parameter $\omega$ arbitrary close to $-1$, the
volume integral quantifier would by itself become infinitesimally
small, independently of the value of the matching radius $a$,
although for this case the wormhole would flare-out very slowly.
Now taking the limit $a\rightarrow r_0$, one verifies that
$I_V\rightarrow 0$. Therefore, as in the examples presented in
Refs. \cite{phantomWH,EOSaccel}, one verifies that, in principle,
one may construct conformally symmetric phantom wormholes with
vanishingly small amounts of phantom energy violating the averaged
null energy condition.

\subsubsection{Tidal acceleration restrictions}

An interesting constraint on the wormhole dimensions, in
particular, on the throat radius may be inferred from the tidal
acceleration restrictions \cite{Morris:1988cz}. The latter
constraints as measured by a traveler moving radially through the
wormhole, are given by the following inequalities
\begin{eqnarray}
\left |\left (1-\frac{b}{r} \right ) \left [\Phi ''+(\Phi ')^2-
\frac{b'r-b}{2r(r-b)}\Phi' \right] \right
|\,\big|\eta^{\hat{1}'}\big| \leq  g_\oplus   \,,
    \label{radialtidalconstraint}
\end{eqnarray}
\begin{eqnarray}
\left | \frac{\gamma ^2}{2r^2} \left [v^2\left (b'-\frac{b}{r}
\right )+2(r-b)\Phi ' \right] \right | \,\big|\eta^{\hat{2}'}\big|
\leq   g_\oplus    \,, \label{lateraltidalconstraint}
\end{eqnarray}
where $\eta^{\hat{i}'}$ is the separation between two arbitrary
parts of his body measured in the traveler's reference frame. We
shall consider $|\eta^{\hat{i}'}|=|\eta|$, for simplicity. We
refer the reader to Ref. \cite{Morris:1988cz} for details. The
radial tidal constraint, inequality (\ref{radialtidalconstraint}),
constrains the redshift function; and the lateral tidal
constraint, inequality (\ref{lateraltidalconstraint}), constrains
the velocity with which observers traverse the wormhole. These
inequalities are particularly simple at the throat, $r_0$,
\begin{eqnarray}
|\Phi '(r_0)| &\leq & \frac{2g_{\oplus}\,r_0}{(1-b')\,|\eta|} \,,
     \label{radialtidalconstraint2}     \\
\gamma^2 v^2 &\leq & \frac{2g_{\oplus}\,r_0^2}{(1-b')\,|\eta|}
\label{lateraltidalconstraint2}  \,.
\end{eqnarray}

From the radial tidal condition (\ref{radialtidalconstraint2}) one
verifies
\begin{equation}
r_0^2 \geq
\left(\frac{1+\omega}{\omega}\right)\frac{|\eta|}{2g_{\oplus}}\,.
    \label{traversal2a}
\end{equation}
Now, considering the equality case for simplicity, assuming that
$|\eta|\approx 2\,{\rm m}$ along any spatial direction in the
traveler's reference frame, and inserting $c$ for clarity, one
verifies that $r_0\simeq c[(1+\omega)/(10\omega)]^{1/2}$. Note
that one may obtain an arbitrary small wormhole throat radius by
imposing that $\omega\rightarrow -1$.

Analogously, from the lateral tidal condition
(\ref{lateraltidalconstraint2}), and considering non-relativistic
velocities, i.e., $\gamma \approx 1$, one has
\begin{equation}
v^2 \leq
\left(\frac{\omega}{1+\omega}\right)\frac{2r_0^2g_{\oplus}}{|\eta|}\,.
    \label{traversal2b}
\end{equation}
Considering the equality case in Eq. (\ref{traversal2a}), one
immediately verifies the following consistency relationship, $v
\leq 1$.

\section{Summary and conclusion}\label{Sec4:Conclusion}

The conventional manner of finding wormhole solutions is
essentially to consider an interesting spacetime metric, and then
deduce the stress-energy tensor components. In this work, we have
considered a more systematic approach in searching for exact
solutions, namely, by assuming spherical symmetry and the
existence of a {\it non-static} conformal symmetry. A wide variety
of solutions with the exotic matter restricted to the throat
neighborhood and with a cut-off of the stress-energy tensor at a
junction interface were deduced, and particular asymptotically
flat geometries were also found. The specific solutions were
deduced by considering choices for the form function, an equation
of state relating the energy density and the anisotropy, and
phantom wormhole geometries were also explored.

Although the assumption of a static conformal symmetry, i.e., with
a static vector $\xi$, was found responsible for the singular
solutions at the center, we emphasize that this is not problematic
to wormhole physics, due to the absence of a center. A wide
variety of the solutions found in this work were considered by
choosing a non-static conformal symmetry, i.e., with a non-static
$\xi$ and static $\psi$. Note that this analysis could be
generalized  by imposing a non-static conformal function
$\psi(r,t)$, where a wider variety of exact solutions may be
found. However, this shall be analyzed in a future work.

\acknowledgments The work of CGB was supported by research grant
BO 2530/1-1 of the German Research Foundation (DFG). The work of
TH was supported by the RGC grant No.~7027/06P of the government
of the Hong Kong SAR. FSNL was funded by Funda\c{c}\~{a}o para a
Ci\^{e}ncia e a Tecnologia (FCT)--Portugal through the grant
SFRH/BPD/26269/2006.


\begin{thebibliography}{99}

\bibitem{Morris:1988cz}
  M.~S.~Morris and K.~S.~Thorne,
  ``Wormholes in space-time and their use for interstellar travel: A tool for
  teaching general relativity,''
  Am.\ J.\ Phys.\  {\bf 56}, 395 (1988).

\bibitem{Lobo:2004wq}
  F.~S.~N.~Lobo and M.~Visser,
  ``Fundamental limitations on `warp drive' spacetimes,''
  Class.\ Quant.\ Grav.\  {\bf 21}, 5871 (2004)
  [arXiv:gr-qc/0406083].

\bibitem{Morris:1988tu}
  M.~S.~Morris, K.~S.~Thorne and U.~Yurtsever,
  ``Wormholes, Time Machines, and the Weak Energy Condition,''
  Phys.\ Rev.\ Lett.\  {\bf 61}, 1446 (1988).

\bibitem{Visser}
  M.~Visser,
  {\it Loretzian wormholes: from Einstein to Hawking}
  AIP Press (1995).

\bibitem{hawkingellis}
  S.~W.~Hawking and G.~F.~R.~Ellis,
  {\it The Large Scale Structure of Spacetime},
  (Cambridge University Press, Cambridge 1973).

\bibitem{WHscalar}
  C.~Barcelo and M.~Visser, ``Traversable wormholes from massless
  conformally coupled scalar fields,'' Phys.\ Lett.\ {\bf B466},
  127-134 (1999) [arXiv:gr-qc/9908029];

  S.~V.~Sushkov and S.-W.~Kim, ``Wormholes supported by a kink-like
  configuration of a scalar field,'' Class.\ Quant.\ Grav.\ {\bf 19},
  4909-4921 (2002) [arXiv:gr-qc/0208069].

\bibitem{Kar}
  B.~Bhawal and S.~Kar, ``Lorentzian wormholes in
  Einstein-Gauss-Bonnet theory,'' Phys.\ Rev.\ {\bf D46}, 2464-2468
  (1992);

  M.~Thibeault, C.~Simeone and E.~F.~Eiroa, ``Thin-shell wormholes in
  Einstein-Maxwell theory with a Gauss-Bonnet term,'' Gen. Relativ. Grav. {\bf 38}, 1593
  (2006) [arXiv:gr-qc/0512029];

  G.~Dotti, J.~Oliva, and R.~Troncoso, ``Static wormhole solution
  for higher-dimensional gravity in vacuum,'' Phys.\ Rev.\ {\bf D75},
  024002 (2007) [arXiv:hep-th/0607062].

\bibitem{BraneWH}
  L.~A.~Anchordoqui and S.~E.~P.~Bergliaffa,
  ``Wormhole surgery and cosmology on the brane: The world is not enough,''
  Phys.\ Rev.\ {\bf D62}, 067502 (2000) [arXiv:gr-qc/0001019];

  K.~A.~Bronnikov and S.~W.~Kim,
  ``Possible wormholes in a brane world,''
  Phys.\ Rev.\ {\bf D67}, 064027 (2003)
  [arXiv:gr-qc/0212112];

  M.~La Camera,
  ``Wormhole solutions in the Randall-Sundrum scenario,''
  Phys.\ Lett.\ {\bf B573}, 27-32 (2003)
  [arXiv:gr-qc/0306017];

  F.~S.~N.~Lobo,
  ``General class of braneworld wormholes,''
  Phys.\ Rev.\ {\bf D75}, 064027 (2007)
  [arXiv:gr-qc/0701133].

\bibitem{Nandi}
  K.~K.~Nandi, B.~Bhattacharjee, S.~M.~K.~Alam, and J.~Evans,
  ``Brans-Dicke wormholes in the Jordan and Einstein frames,''
  Phys.\ Rev.\ {\bf D57}, 823-828 (1998).

\bibitem{Garattini:2007ff}
  R.~Garattini and F.~S.~N.~Lobo,
  ``Self sustained phantom wormholes in semi-classical gravity,''
  Class.\ Quant.\ Grav.\ {\bf 24}, 2401 (2007)
  [arXiv:gr-qc/0701020].

\bibitem{Arellano}
  K.~A.~Bronnikov,
  ``Regular magnetic black holes and monopoles from nonlinear
  electrodynamics,''
  Phys.\ Rev.\ {\bf D63}, 044005 (2001)
  [arXiv:gr-qc/0006014];

  A.~V.~B.~Arellano and F.~S.~N.~Lobo,
  ``Evolving wormhole geometries within nonlinear electrodynamics,''
  Class.\ Quant.\ Grav.\ {\bf 23}, 5811 (2006)
  [arXiv:gr-qc/0608003];

  A.~V.~B.~Arellano and F.~S.~N.~Lobo,
  ``Non-existence of static, spherically symmetric and stationary, axisymmetric
  traversable wormholes coupled to nonlinear electrodynamics,''
  Class.\ Quant.\ Grav.\ {\bf 23}, 7229 (2006)
  [arXiv:gr-qc/0604095].

\bibitem{phantomWH}
  F.~S.~N.~Lobo,
  ``Phantom energy traversable wormholes,''
  Phys.\ Rev.\ {\bf D71}, 084011 (2005)
  [arXiv:gr-qc/0502099];

  F.~S.~N.~Lobo,
  ``Stability of phantom wormholes,''
  Phys.\ Rev.\ {\bf D71}, 124022 (2005)
  [arXiv:gr-qc/0506001].

\bibitem{phantomWH2}
  S.~Sushkov, ``Wormholes supported by a phantom energy,''
  Phys.\ Rev.\ {\bf D71}, 043520 (2005) [arXiv:gr-qc/0502084];

  O.~B.~Zaslavskii, ``Exactly solvable model of wormhole supported
  by phantom energy,'' Phys.\ Rev.\ {\bf D72}, 061303 (2005)
  [arXiv:gr-qc/0508057].

\bibitem{EOSaccel}
  F.~S.~N.~Lobo,
  ``Chaplygin traversable wormholes,''
  Phys.\ Rev.\ {\bf D73}, 064028 (2006)
  [arXiv:gr-qc/0511003];

  F.~S.~N.~Lobo,
  ``Van der Waals quintessence stars,''
  Phys.\ Rev.\ {\bf D75}, 024023 (2007)
  [arXiv:gr-qc/0610118];

  J.~A.~J.~Madrid, ``Chaplygin gas may prevent big trip,''
  Phys.\ Lett.\ {\bf B634} 106 (2006) [arXiv:astro-ph/0512117];

  E.~F.~Eiroa and C.~Simeone, ``Stability of Chaplygin gas thin-shell
  wormholes,'' Phys. Rev. {\bf D76}, 024021 (2007) [arXiv:0704.1136].

\bibitem{gonzalez}
  P.~F.~Gonz\'alez-D\'{\i}az, ``Wormholes and ringholes in a
  dark-energy universe,'' Phys.\ Rev.\ {\bf D68}, 084016 (2003)
  [arXiv:astro-ph/0308382];

  P.~F.~Gonz\'{a}lez-D\'{i}az, ``Achronal cosmic future,''
  Phys.\ Rev.\ Lett.\ {\bf 93} 071301 (2004) [arXiv:astro-ph/0404045];

  P.~F.~Gonz\'{a}lez-D\'{i}az and J.~A.~J.~Madrid, ``Phantom
  inflation and the `Big Trip','' Phys.\ Lett.\ {\bf B596} 16-25
  (2004) [arXiv:hep-th/0406261];

  P.~F.~Gonz\'{a}lez-D\'{i}az,
  ``On the accretion of phantom energy onto wormholes,''
  Phys.\ Lett.\ {\bf B632}, 159 (2006)
  [arXiv:astro-ph/0510771].

\bibitem{Lemos:2003jb}
  J.~P.~S.~Lemos, F.~S.~N.~Lobo and S.~Quinet de Oliveira,
  ``Morris-Thorne wormholes with a cosmological constant,''
  Phys.\ Rev.\ {\bf D68}, 064004 (2003)
  [arXiv:gr-qc/0302049].

\bibitem{Herrera}
  L.~Herrera, J.~Jimenez, L.~Leal, J.~Ponce de Leon, M.~Esculpi and
  V.~Galina, ''Anisotropic fluids and conformal motions in general
  relativity,'' J.\ Math.\ Phys.\ {\bf 25}, 3274 (1984);

  L.~Herrera and J.~Ponce de Leon, ''Isotropic and anisotropic charged spheres
  admitting a one-parameter group of conformal motions,'' J.\ Math.\ Phys.\ {\bf 26},
  2302 (1985).

\bibitem{Maartens:1989ay}
  R.~Maartens and M.~S.~Maharaj,
  ``Conformally symmetric static fluid spheres,''
  J.\ Math.\ Phys.\ {\bf 31}, 151 (1990).

\bibitem{confpf}
  H.~Stephani,
  ``{\"U}ber {L}{\"o}sungen der {E}insteinschen {F}eldgleichungen, die
  sich in einen f{\"u}nfdimensionalen flachen {R}aum einbetten lassen,''
  Commun.\ Math.\ Phys.\ {\bf 4}, 137 (1967);

  H.~Stephani, D.~Kramer, M.~MacCallum, C.~Hoenselaers, and E.~Herlt,
 {\it Exact solutions of Einstein's field equations},
 Cambridge University Press (2003);

 C.~G.~B\"{o}hmer,
 ``General Relativistic Static Fluid Solutions with Cosmological Constant,''
 [arXiv:gr-qc/0308057], {\it unpublished Diploma Thesis}, (2003).

\bibitem{Mak:2003kw}
  M.~K.~Mak and T.~Harko,
  ``Quark stars admitting a one-parameter group of conformal motions,''
  Int.\ J.\ Mod.\ Phys.\ {\bf D13}, 149 (2004)
  [arXiv:gr-qc/0309069].

\bibitem{Harko:2004ui}
  T.~Harko and M.~K.~Mak,
  ``Vacuum solutions of the gravitational field equations in the brane world
  model,''
  Phys.\ Rev.\ {\bf D69}, 064020 (2004)
  [arXiv:gr-qc/0401049].

\bibitem{Mak:2004hv}
  M.~K.~Mak and T.~Harko,
  ``Can the galactic rotation curves be explained in brane world models?,''
  Phys.\ Rev.\ {\bf D70}, 024010 (2004)
  [arXiv:gr-qc/0404104].

\bibitem{Darmois-Israel}
  W.~Israel, ``Singular hypersurfaces and thin shells in general
  relativity,''   Nuovo Cimento {\bf B44}, 1 (1966); and corrections
  in {\it ibid.} {\bf B48}, 463 (1966).

\bibitem{LoboCrawford}
  F.~S.~N.~Lobo and P.~Crawford, ``Stability analysis of dynamic
  thin shells,'' Class.\ Quant.\ Grav.\ {\bf 22}, 4869 (2005),
  [arXiv:gr-qc/0507063].

\bibitem{wormhole-shell}

 E.~Poisson and M.~Visser, ``Thin-shell wormholes: Linearization
 stability,'' Phys. Rev. {\bf D52}, 7318 (1995) [arXiv:gr-qc/9506083];

F.~S.~N.~Lobo and P.~Crawford, ``Linearized stability analysis of
 thin-shell wormholes with a cosmological constant,''
 Class.\ Quant.\ Grav.\ {\bf 21}, 391 (2004) [arXiv:gr-qc/0311002];

  E.~F.~Eiroa and G.~E.~Romero, ``Linearized stability of charged thin-shell
  wormholes,'' Gen. Relativ. Grav. {\bf 36}, 651 (2004)
  [arXiv:gr-qc/0303093];

  J.~P.~S.~Lemos and F.~S.~N.~Lobo, ``Plane symmetric traversable
  wormholes in an anti-de Sitter background,'' Phys.\ Rev.\ {\bf D69},
  104007 (2004) [arXiv:gr-qc/0402099];

  F.~S.~N.~Lobo, ``Surface stresses on a thin shell surrounding a
  traversable wormhole,'' Class. Quant. Grav. {\bf 21}, 4811 (2004)
  [arXiv:gr-qc/0409018];

  F.~S.~N.~Lobo, ``Energy conditions, traversable wormholes and dust
  shells,'' Gen. Rel. Grav. {\bf 37}, 2023-2038 (2005)
  [arXiv:gr-qc/0410087].

\bibitem{foot1}
  The Lambert W function is implicitly defined by the following relation
  ${\cal W}(x) e^{{\cal W}(x)} = x$.

\bibitem{Visser:2003yf}
  M.~Visser, S.~Kar and N.~Dadhich,
  ``Traversable wormholes with arbitrarily small energy condition violations,''
  Phys.\ Rev.\ Lett.\  {\bf 90}, 201102 (2003)
  [arXiv:gr-qc/0301003].

\end{thebibliography}
\end{document}